\documentclass[sigconf]{acmart}
\AtBeginDocument{%
  }

\usepackage{hyperref}
\usepackage{boldline}
\usepackage{xcolor}
\usepackage{tabularx}
\usepackage{tcolorbox}
\usepackage{color, colortbl}
\usepackage{amsmath,amsfonts}
\usepackage{amsmath}
\usepackage{array}
\usepackage{algorithmic}
\usepackage{balance}
\usepackage{blindtext}
\usepackage{multirow}
\usepackage{caption}
\usepackage{hyperref}
\usepackage{subcaption}
\usepackage[compact]{titlesec}
\usepackage{enumitem}
\setlist{nosep}
\usepackage[skip=2pt]{caption}
\usepackage{parskip}

\begin{document}

\newcommand{\sarah}[1]{\ifshowcomments{{\color{blue}Sarah: #1}}\fi}
\newcommand{\shraddha}[1]{\ifshowcomments{{\color{brown}Shraddha: #1}}\fi}
\definecolor{mySpec}      {RGB}{255,199,  0}   
\definecolor{myProof}     {RGB}{163, 73,164}   
\definecolor{myPaused}    {gray}{0.50}         
\definecolor{myImpl}      {RGB}{237,125, 49}   
\definecolor{myTest}      {RGB}{128, 64,  0}   
\definecolor{myComment}   {RGB}{120, 60,  0}   
\definecolor{myLookup}    {RGB}{  0,128,128}   
\definecolor{myStructure} {RGB}{ 68,114,196}   
\definecolor{myVerify}    {RGB}{200,  0,  0}   
\definecolor{myReadErr}   {RGB}{  0,  0,  0}   

\newcommand{\stateSpec}{\textcolor{mySpec}{\rule{6pt}{6pt}}}
\newcommand{\stateProof}{\textcolor{myProof}{\rule{6pt}{6pt}}}
\newcommand{\stateImpl}{\textcolor{myImpl}{\rule{6pt}{6pt}}}
\newcommand{\statePaused}{\textcolor{myPaused}{\rule{6pt}{6pt}}}

\newcommand{\verifyDot}{\textcolor{myVerify}{\ding{217}}}   
\newcommand{\readerrDot}{\textcolor{myReadErr}{\ding{217}}}

\title{What's in a Proof? Analyzing Expert Proof-Writing Processes in F* and Verus}



\author{Rijul Jain}
\email{rijul@cs.washington.edu}
\affiliation{%
  \institution{University of Washington}
  \city{Seattle}\state{WA}  \country{USA}
}

\author{Shraddha Barke }
\email{sbarke@microsoft.com}
\affiliation{%
  \institution{Microsoft Research}
  \city{Redmond}\state{WA}  \country{USA}
}

\author{Gabriel Ebner }
\email{gabrielebner@microsoft.com}
\affiliation{%
  \institution{Microsoft Research}
  \city{Redmond}\state{WA}  \country{USA}
}

\author{Md Rakib Hossain Misu }
\email{mdrh@uci.edu>}
\affiliation{%
  \institution{University of California Irvine}
  \city{Irvine}\state{CA}  \country{USA}
}

\author{Shan Lu}
\email{shanlu@microsoft.com}
\affiliation{%
  \institution{Microsoft Research}
  \city{Redmond}\state{WA}  \country{USA}
}

\author{Sarah Fakhoury}
\email{sfakhoury@microsoft.com}
\affiliation{%
  \institution{Microsoft Research}
  \city{Redmond}\state{WA}  \country{USA}
}

\renewcommand{\shortauthors}{Jain et al.}

\begin{abstract}
    Proof-oriented programming languages (POPLs) empower developers to write code alongside formal correctness proofs, providing formal guarantees that the code adheres to specified requirements.  
Despite their powerful capabilities, POPLs present a steep learning curve and
have not yet been adopted by the broader software community. The lack of understanding about
the proof-development process and how expert proof developers interact with POPLs has 
hindered the advancement of effective proof engineering and the development of proof-synthesis models/tools.  
In this work, we conduct a user study, involving the collection and analysis of fine-grained source code telemetry from eight experts working with two languages, F* and Verus. Results reveal interesting trends and patterns about 
how experts reason about proofs and key challenges encountered during the proof development process. We identify three distinct strategies and multiple informal practices 
that are not captured final code snapshots, yet are predictive of task
outcomes. We translate these findings into concrete design guidance for
AI proof assistants: bias toward early specification drafting, explicit sub‑goal
decomposition, bounded active errors, and disciplined verifier
interaction.  We also present a case study of an F* proof agent grounded in these recommendations, and demonstrate improved performance over baseline LLMs. 
\end{abstract}

\maketitle

\section{INTRODUCTION}
\begin{figure*}[h!]
  \centering
  \begin{minipage}{\textwidth}
    \includegraphics[width=1\linewidth]{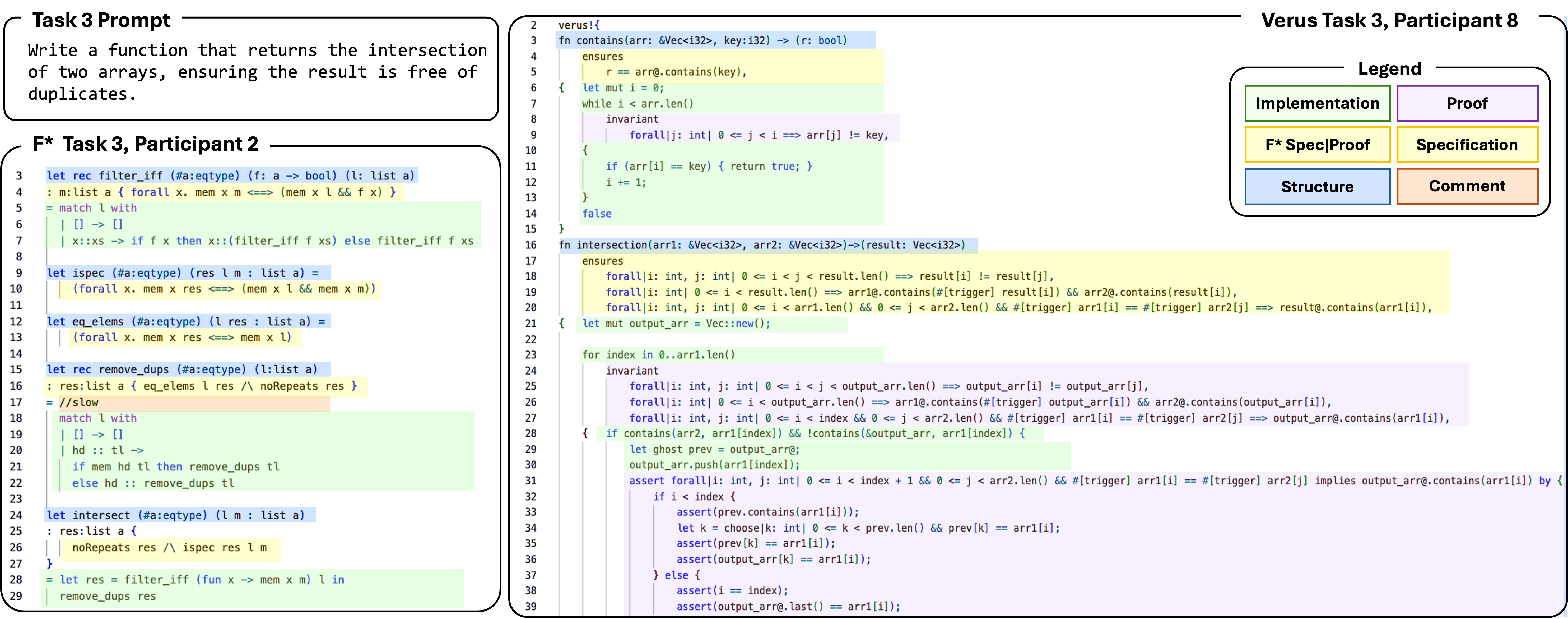}
  \end{minipage}
  \caption{Two solutions for Task 3 in F* (left) and Verus (right). Each line of code is colored according to the state it belongs to. }
  \label{intro}
\end{figure*}
\emph{Proof‑oriented programming languages} (POPLs) such as
F* and Verus weave formal logic into ordinary source code:
alongside an implementation, the developer writes specifications,
loop invariants, and auxiliary lemmas;
an SMT‑based verifier then proves, or rejects, the program’s
correctness before it ever runs. POPLs enable scalable high-assurance programs \cite{highassurance} for critical systems from privacy-preserving networking protocols \cite{protocols} to confidential virtual machines 

Despite their powerful capabilities, using POPLs and producing proofs remains an high-barrier skill,
learning curves are steep, solver feedback is often cryptic, and even experts
must engage in trial‑and‑error loops that hinder productivity~\cite{untangling}. Recent work hints that large language models (LLMs) could automate sizeable fragments of the proof burden, either by synthesizing verified code end‑to‑end or by iteratively repairing failed proofs ~\cite{aggarwal2024alphaverusbootstrappingformallyverified,neural, daf, baldur}. 

However, unlike the code-writing process for mainstream languages~\cite{setextbook}, most aspects of the proof-writing process, such as reasoning strategies and IDE interactions, remain largely understudied. A ground‑truth picture of how experts
actually engineer proofs is missing:  Which edits dominate their time? How do they orchestrate verifier calls and manual reasoning? Do distinct expert \emph{strategies} exist that we could imitate or enforce in automated agents? Answering these questions is a prerequisite for both principled tool design and data‑efficient training of LLM proof assistants.

This paper presents a user study of how experts approach the proof-writing process in two languages, F$^*$ \cite{mumon} and Verus \cite{verus}. We instrumented the VSCode extensions for F* and Verus to record
fine‑grained telemetry while eight experts POPL users solved four
benchmark tasks. Our telemetry dataset comprises over 18,000+ telemetry events including: every code edit, verifier
invocation, active errors, pause states, hover lookup, and comment toggle. We automatically annotate each event with its proof‑writing
state (e.g.\ \textsc{Spec}, \textsc{Impl}, \textsc{Proof}) and action categories.  

Results reveals three distinct strategies: \textsc{Spec‑first
Planners}, \textsc{Rapid Verifiers}, and a \textsc{balanced} strategy and 
link the \textsc{Spec-first Planner} style to both faster completion and higher task success rates. Across both languages, an effective proof strategy is characterized
by (i) deliberate early specification drafting, (ii) measured
verifier invocation, and (iii) disciplined error management with frequent
returns to a clean state, i.e. where no active errors are present in the file.
Furthermore, to explore actionable value, we instantiate a
\textbf{proof agent} that is  conditioned on the insights of this work.
The agent design aims to mimic the specification-first planner strategy and preliminary evaluation shows performance boost over baseline LLMs.  \textbf{This work makes the following contributions:}
\begin{enumerate}[leftmargin=1.75em]
  \item A  dataset of 18k+ annotated telemetry
        events from expert F* and Verus sessions.
  \item The first quantitative breakdown of where F* and Verus proof engineers
        invest effort and how they  interact with the verifier.
  \item Identification of three distinct strategy archetypes via
        K‑means clustering, backed by timeline case studies and results of thematic analysis on self-reported strategies.
  \item A proof‑of‑concept LLM‑driven proof agent that embeds these findings shows promising performance against baseline LLMs.
\end{enumerate}

Our results show that seemingly informal activities like early
specification drafting, deliberate pauses, and disciplined engagement with the verifier, are indicative hallmarks that lead to effectively solving proof tasks.
We argue that future POPL tooling and LLM proof assistants should be informed by these strategies.


\section{BACKGROUND}

\subsection{Proof-oriented Programming Languages: Verus and F*}

\textbf{Verus} is a novel state-of-the-art verification language designed to extend code written in Rust \cite{verus}. For every function to be proved, users express the intended function behavior through pre- and postconditions in corresponding \texttt{requires or ensures} specification code blocks (the yellow portion of \autoref{intro}), and provide proof code (highlighted in purple) in the form of loop invariants when required by the underlying verifier. 
Verus then composes queries for the underlying SMT solver, querying whether there exists any value to the function arguments that can satisfy the precondition and cause the postcondition to fail. If the solver answers `no', the proof is done; if the solver answers `yes', either the code is buggy or the solver needs extra hints to accomplish the proof. 

\textbf{F*} is a general-purpose proof-oriented programming language~\cite{mumon} that has been widely used in high-assurance software projects such as the Linux kernel, Mozilla Firefox, and Windows Hyper-V \cite{haclstar,everparse3d}. F* is
similar to Verus in using an SMT solver~\cite{de2008z3} to verify the proof.  F* differs from Verus in several ways:
1) Specification is mainly expressed through dependent types instead of inside special code blocks, using refinement types to annotate input and output types with pre- and postconditions.
2) This leads to an \emph{intrinsic} proof style, where code and proofs are intermixed. In many cases, the postconditions provided by the function calls are already sufficient and no extra proof steps are needed. When extra proof steps are required, this typically means the invocation of a lemma. 3) The idiomatic style encouraged by F* is functional programming, making use of inductive datatypes and structural recursion, as opposed to Verus, where it is common to use arrays, for example.

\subsection{User studies for Proof Development}
Prior studies on proof development relied on existing datasets in the form of source code files~\cite{aspinall2016towards}, project logs and version control history~\cite{andronick2012large, staples2014productivity},  Isabelle archives~\cite{blanchette2015mining} and social channels of popular proof assistants like Lean, Coq, and Isabelle~\cite{lincroft2024thirty}.
This work is one of the first to conduct a multilingual user study of POPL experts by collecting fine-grained user interaction telemetry data and survey responses.
%
%
\cite{ringer2020replica} conducted a detailed month-long Coq study by instrumenting Coq’s interaction model to collect fine-grained data on uncovering patterns in proof developments and interactions.
Our findings echo \cite{ringer2020replica}, who observed that more than 75\% of the proof attempts required rewriting initial definitions or specifications to allow successful proof writing.
We also observed frequent refinement of the specifications throughout the task to align them with the implementation.
Recently, Shi et al.~\cite{shi2025qed} conducted an in-depth contextual inquiry of 30 proof assistant users working with Rocq and Lean.
They identified several themes similar to ours: proof writing is shaped by iterative feedback loops with the assistant, often resembles an interactive process, and relies heavily on external resources.
Building on these observations, we take a step further by proposing an agentic system case study design, grounded in insights from our expert user study.
\citet{tavante2021data} observed the critical role of user interfaces in proof assistant tools and the steep learning curve associated with the Coq development environment for new users. 
In our study, even experts found it a challenge to work in the study environment without their internal tools setup.
%
%
%
%
%
%
%

\subsection{Language Models for Proof Generation}
Nascent work to advance language models capabilities for automated theorem proving is inspired by human reasoning steps.
Lean-Star~\cite{lin2024lean} generates synthetic natural language "thoughts" during training, interleaving intuitive insights with formal tactics.
Lean Copilot \cite{song2024towards} provides step-specific tactic suggestions and searches for multi-tactic proofs by combining LLM-generated proof steps with Lean's existing rule-based proof search.
%
LLMStep~\cite{welleck2023llmstep} operates step-by-step, querying an LLM with the current proof state to generate and validate proof suggestions within Lean. 
 Our results indicate that planning ahead, frequent and iterative reliance on the verifier, as well as task decomposition may be additional signals critical to informing future models.  
%

\section{RESEARCH QUESTIONS}

To understand how experts actually invest their time and attention when using proof-oriented programming languages, we focus on three dimensions: (i) \emph{where} effort accrues across the proof writing process, (ii) \emph{how} experts interact with the verifier, and (iii) \emph{which} higher‑level proof engineering strategies they adopt and how those strategies impact measurable task outcomes. Accordingly we ask:

\begin{itemize}[leftmargin=1.75em]
  \item[RQ1] \textbf{How is expert effort distributed across proof states (specification, implementation, proof) in F* and Verus?}

  \item[RQ2] \textbf{What verifier interaction patterns (frequency, timing) do experts employ?}

  \item[RQ3] \textbf{Which proof engineering strategies do experts employ, and how do these strategies relate to task outcomes?}

  \begin{itemize}[leftmargin=2.0em, nosep]
      \item[RQ3a] \textbf{What strategy themes emerge from a thematic analysis of post‑task questionnaires?}

      \item[RQ3b] \textbf{What data‑driven strategy archetypes arise from telemetry, and do they predict task success and efficiency?}
    \end{itemize}
\end{itemize}

\noindent Collectively, answering RQ1–RQ3 provides a layered analysis of: low‑level time allocation, mid‑level interaction dynamics, and high‑level strategies of expert proof engineering practice across two verification languages. Finally, grounded in these results, we propose a proof-agent design and evaluate against a baseline LLM on a subset of the tasks used in this study.

\section{USER STUDY}

To investigate expert proof-writing processes, we conduct a user study in which we collect live telemetry data during various proof tasks across two proof-oriented languages, F* and Verus, in a controlled GitHub Codespace environment.

\textbf{IDE plugins} for VSCode have been developed for both Verus and F*. These plugins offer light-weight syntax checks and easy ways to invoke verification and view errors when verification fails. For the purpose of our study, we extend both of the existing F* and Verus plugins to collect telemetry data. We do this by adding hooks to all calls to the Language Server, edits, and actions taken in the VSCode editor and store this data into a database.  
For both F* and Verus, any code-save action (e.g., `Ctrl-S') will trigger a scan of the code and a query of the underlying SMT solver. 
In the F* plugin, there is also a command to verify the code up until the cursor. The verification results are displayed inside VS Code: when the verification fails, corresponding code snippets that cannot be verified will be highlighted. When the verification succeeds, a pop-up message box will say so in Verus. 
The F* extension shows the verification status using a bar on the left side of the editor; verified portions of the file are shown in green.

\subsection{Tasks}
The study comprises of four proof tasks, detailed in \autoref{tab:tasks-table}. 
Task 1 is a common task that appears in the tutorials of both Verus and F*. Tasks 2--4 are selected from the MBPP~\cite{austin2021program} benchmark suite. 
We select tasks based on these criteria:
1) the original description of the coding task has no ambiguity while also leaving some flexibility in data structure choices; 
2) the proof task should have medium difficulty --- complex enough to reveal challenges in proof-writing but not overly fatiguing.
3) the approaches used to solve the four tasks should be sufficiently diverse.
To ensure validity, we recruited an expert for each language, excluded from the study, who helped refine tasks and ground-truth solutions.
A pilot study was conducted prior to data collection to validate task difficulty, telemetry instrumentation, and study procedures.

\begin{table*}[]
\resizebox{\linewidth}{!}{%
\begin{tabular}{lllllll}
\hline
Task & Task Name & Description & Language & Participants & Duration & Effort \\ \hline
\multirow{2}{*}{T1} & \multirow{2}{*}{\sc BinarySearch} & \multirow{2}{*}{\begin{tabular}[c]{@{}l@{}}Implementation of binary search\end{tabular}} & F* & P1, P4 & \bf{45.07} & \bf{7.5/10} \\
 &  &  & Verus & P6, P7 & 11.80 & 2/10 \\ \hline
\multirow{2}{*}{T2} & \multirow{2}{*}{\sc Sublist} & \multirow{2}{*}{\begin{tabular}[c]{@{}l@{}}Checks whether a list is a \\ contiguous sublist of another or not\end{tabular}} & F* & P2, P3 & 37.06 & 6/10 \\
 &  &  & Verus & P5, P8 & 35.35 & 3.5/10 \\ \hline
\multirow{2}{*}{T3} & \multirow{2}{*}{\sc Intersection} & \multirow{2}{*}{\begin{tabular}[c]{@{}l@{}}Returns the intersection of two \\ arrays ensuring the result is free of duplicates.\end{tabular}} & F* &  P1, P2, P4 & 16.70 & 5/10 \\
 &  &  & Verus & P7, P8 & 34.40 & 5.5/10 \\ \hline
\multirow{2}{*}{T4} & \multirow{2}{*}{\sc MaxDifference} & \multirow{2}{*}{\begin{tabular}[c]{@{}l@{}}Returns the difference between the\\  largest and smallest value in a given list.\end{tabular}} & F* & P1, P3 & 14.17 & 4.5/10 \\
 &  &  & Verus & P5, P6 & 19.00 & 3/10 \\ \hline
\end{tabular}%
}
\caption{Four tasks included in the user study, including a brief description of the task. Participants assigned to each task and average task duration (minutes) within each language, and average self-reported effort (scale 0-10).  }
\label{tab:tasks-table}
\end{table*}

\subsection{Participants}

We recruited eight participants from Microsoft Research, a large trans-national industry research lab, all of whom were experts in either F* or Verus. Six of the eight participants are maintainers or designers of the target languages, while the remaining two (one in F* and one in Verus) have intermediate proficiency in the target languages. Participants were evenly distributed between the two languages, with four participants each assigned to F* and Verus tasks.

\subsection{Procedure}

Participants were randomly assigned two tasks, ensuring each task was completed by two F* and two Verus participants. One F* expert completed a third task to keep tasks balanced; however, results are consistent with the session omitted. 
Each study session was designed to last approximately 90 minutes and was conducted in a pre-configured GitHub Codespace environment. 
Each Codespace was set up with the necessary tools for the assigned language as well as two task files. Codespaces were configured to include our extension of the Verus and F* plugins specifically developed to continuously collect telemetry data.
Each proof task file included 1) a natural language description of the implementation (listed in \autoref{tab:tasks-table}; 2) test cases the implementation should pass; 3) for Verus, we also included a correct Rust implementation that we developed, since our goal is to study the proof-writing process. For F*, the implementation and the proof are more tightly intertwined, making them harder to separate.

After completing each task, participants filled out a structured post-task survey to provide subjective insights into their proof writing process. 

\subsection{Measures}

We collect both quantitative and qualitative measures:
\begin{enumerate}

\item \textbf{Telemetry Data}: Comprehensive edit logs, including timestamps, source code edits, VSCode actions, and timestamp and output of every verifier invocation, resulting in over 18,000+ annotated telemetry events.
\item \textbf{Task Completion Time and Correctness} evaluated based on 1) supplied tests and 2) the verifier.
\item \textbf{Screen Recordings}: Video recordings of participants’ screens, excluding audio, were used to cross-verify automatically annotated telemetry data.
    \item  \textbf{Survey Responses}:  After completing each task, participants filled out a structured post-task survey to provide subjective insights into their proof writing process. They answered long-form questions related to: 1) task difficulty 2) describing the concrete structure of their solution  3) if and how they leveraged the documentation and external libraries 3) a high-level conceptual approach to solve the problem. In addition, they rated task difficulty and their effort. We use the results of this survey and conduct a Thematic Analysis to identify common themes around successful and unsuccessful strategies reported by expert users. We use the results to inform and ground our qualitative analysis of RQ3.  
\end{enumerate}
\subsection{Annotating Telemetry Data}
To answer our research questions, we annotate the raw telemetry data with tags that categorize the types of source code edits being made. 
To achieve this, we first develop a taxonomy of edits and actions that users can perform in both F* and Verus. Next, we identify language-specific keywords that distinguish each type of edit and use these keywords to automatically label each telemetry event, assigning it to a specific stage in the proof-writing process. A source code edit can be one of the following proof-writing states:
\begin{enumerate}
\item \textbf{Specification}: edits related to the program specification (in Verus, pre- and postconditions; in F*, all proof-related edits -- including refinement types

\item \textbf{Proof}: edits related to the proof aside from the specification in Verus, i.e. loop invariants

\item \textbf{Implementation}: edits related to the core implementation of the program functionality

\item \textbf{Structure}: edits made to the overall structure of a solution, including proof structure in F*, e.g. function signatures of proofs or lemmas

\item \textbf{Comment}: edits made in comments, including commenting out source code to defer verification for later

\item \textbf{Test}: edits made to any code relating to a test or testing infrastructure.  In F* for the selected tasks, participants used \texttt{assert} as a form of static unit tests; we therefore annotated \texttt{assert} as test.
\end{enumerate}
In addition to edits in proof states, we process telemetry data to identify user actions such as verification triggers, documentation look up, hover actions, or pauses.

\section{RESULTS: Proof Effort and Verifier Usage } 
At a high level, for F*, 7/8 tasks were successfully completed, while Verus saw 6/8 completions. Unsuccessful attempts included F* (T3; P1) and Verus (T3, P7; T2, P5).
 To answer our research questions, we analyze the telemetry data below.

\begingroup
\footnotesize
\begin{figure*}
    
\centering
\begin{subfigure}{0.4\textwidth}
\centering
\includegraphics[width=\textwidth]{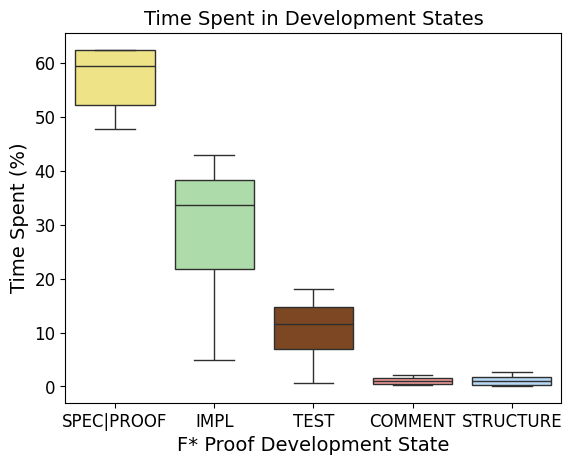}
\label{fstar-time-state}
\end{subfigure}%
\begin{subfigure}{0.4\textwidth}
\centering
\includegraphics[width=\textwidth]{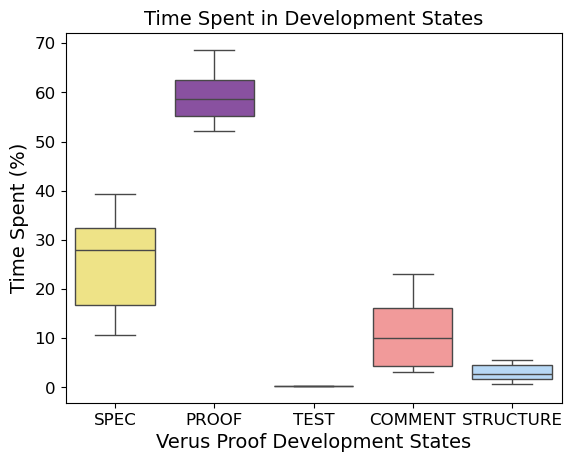}
\label{verus-time-state}
\end{subfigure}

\begin{subfigure}{0.4\textwidth}
\centering
\includegraphics[width=\textwidth]{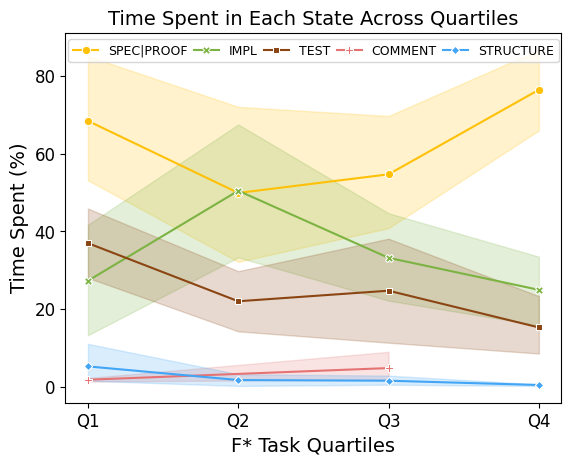}
\label{fstar-time-quartiles}
\end{subfigure}%
\begin{subfigure}{0.4\textwidth}
\centering
\includegraphics[width=\textwidth]{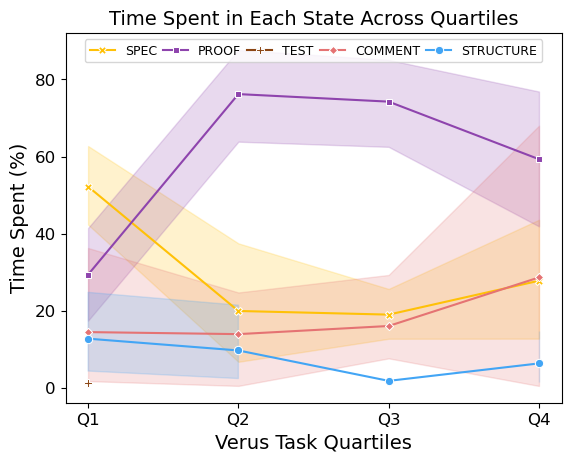}
\label{verus-time-quartiles}
\end{subfigure}
\caption{RQ1: Percentage of time spent in each proof-writing state in F* and Verus, overall and split across task time quartiles. Each task quartile represents 25\% of the task duration. The line plots depict mean trends, with shaded areas reflecting participant variation.}
\label{time}
\end{figure*}
\endgroup

\subsection{RQ1: Where Users Spend Time in the Proof-Writing Process}
To answer RQ1, we first explore the overall proportion of time participants spend in each proof-development state, then we explore how time in each state changes over the course of a task. 

\subsubsection{Proof-Development State Breakdown}
Looking at the overall task durations in \autoref{tab:tasks-table}, we observe similar time spent for each task in both F* and Verus, between 11.8 - 45 minutes on average. There are notable differences for T3 and T1, where F* participants spent almost 34 minutes more on average for T1, and Verus participants spent 9 minutes on average more for T3. For T1, this is the result of the additional time needed to write the implementation in F*. For T3, this is explained by needing to write much more proof code in Verus as compared to F*, highlighted in \autoref{intro}. 
\autoref{time} contains a breakdown of states for each language. For F*, despite the additional effort required to plan and write the implementation of the function to be proven, a majority (55-65\%) of the time was spent writing the specification/proof, with between 20-40\% on the implementation. In addition, for F* tasks, testing the implementation to adhere to the expected behaviors took between 10-20\% of the time, which the Verus tasks did not require since the implementation was provided to the participants. 
For Verus, \autoref{time} shows that the most significant portion of time is spent writing the proof (around 60\%), and the variation in time spent writing proof is minimal regardless of  tasks or participants.  Finally, time spent commenting, or editing comments, is 5-15\% in Verus, more than the 5\% in F*.  Inspecting the telemetry data and screen recordings, we observe comments in Verus are mostly used as a common way to debug and decompose tasks, commenting out part of the code implementation and specification so that it is possible to focus on one proof sub-task at a time. In contrast, in F* users were more likely to defer verification through \texttt{admit} or \texttt{assume} keywords, rather than commenting or editing commented code. 


%
%
%



\subsubsection{Proof-Development Process: How Do States Change Over Time?}

In the bottom two graphs of \autoref{time}, we show the percentage of time spent in each state across task time quartiles.
A key trend is that the development stages of specification, proof, and implementation are not isolated to the start or end of a task, but are intermingled throughout. 
For F*, time spent writing the code implementation and specification/proof is significant across all quartiles. 
Participants typically start by drafting the specification and partially implementing functionality in the first quartile.
Code implementation picks up and reaches its peak during the second quartile. After that, more effort is put in specification development, and yet modification to implementation occurs even during the last quartile. 
%
%
The trend is similar in Verus. 
Despite the absence of code implementation, specification and proof writing are intertwined. 
Specification writing dominates the first quartile but still takes around 10\% of time in the next three quartiles. Proof writing takes the majority of time in quartiles 2 and 3. Particularly, we observed participants edit specifications when they encountered difficulty in writing the proof --- either to fix errors uncovered during proof writing or to simplify the proof process. Unsurprisingly, comment edits peak in the last quartile due to code cleanup.




\subsection{RQ2: Interaction with the Verifier}
To answer RQ2, we explore how experts interact with the verifier by 1) the frequency of verifier invocations at different points throughout the task and 2) how they interact with errors reported by the verifier throughout the task.
\subsubsection{Frequency of Verifier Invocations}
\begin{figure*}[h]
\centering
\begin{subfigure}{0.45\textwidth}
\centering
\includegraphics[width=\textwidth]{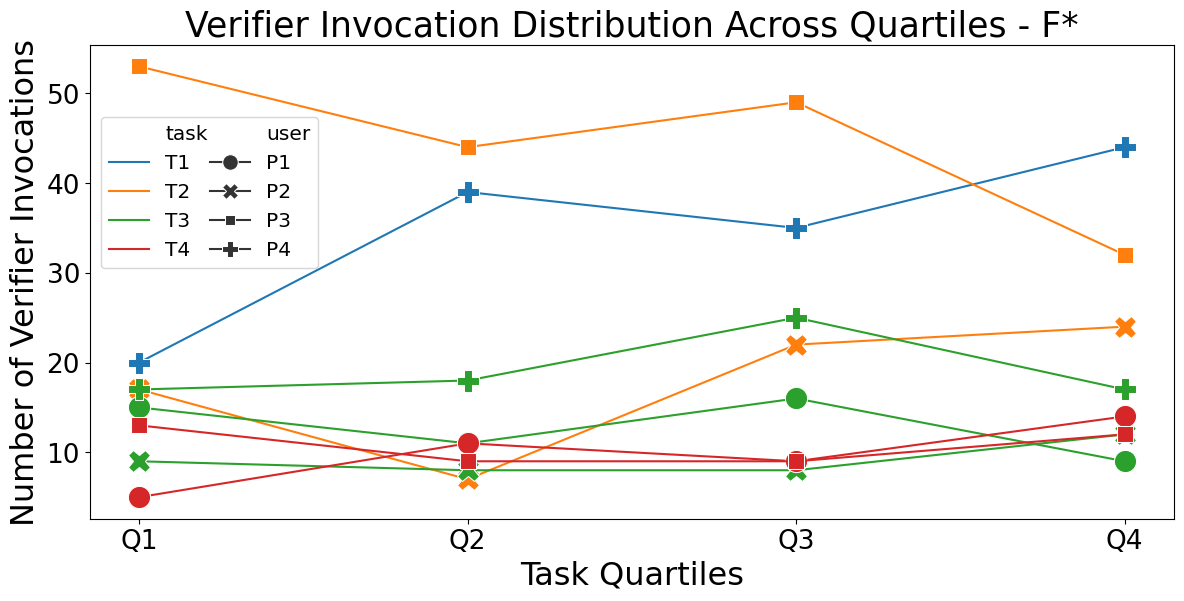}
\label{fstar-verifier}
\end{subfigure}%
\begin{subfigure}{0.47\textwidth}
\centering
\includegraphics[width=\textwidth]{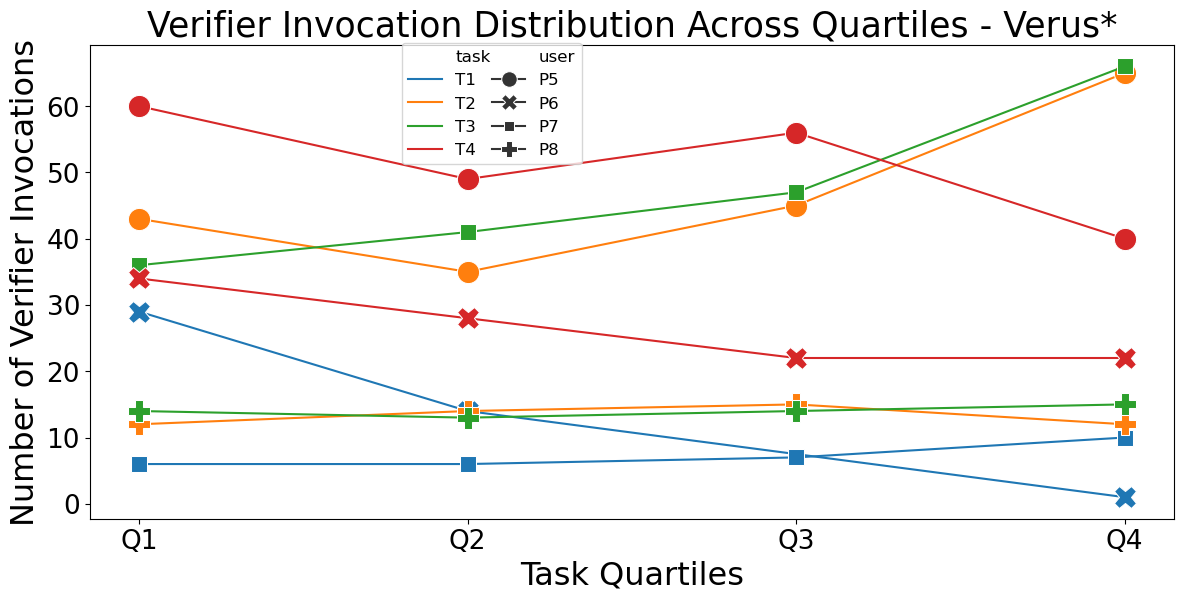}
\label{verus-verifier}
\end{subfigure}
\caption{Number of verifier invocations across task time quartiles for all participants and tasks in F* and Verus, respectively.  }
\label{verifier-invocations}
\end{figure*}
\autoref{verifier-invocations} shows verifier invocation counts across four task quartiles. 
The average rate of verifier invocations across tasks was 3.39 per minute for F* and 9.11 for Verus.
For F*, on the left in \autoref{verifier-invocations}, we see that more difficult tasks show more frequent verifier invocations throughout the task. 
For example, see P3's T2, rated as \textit{`Neither easy nor difficult'} (3/5) but at 6/10 effort, which is relatively high for this task pool, shows between around 30 and 50 invocations in each quartile. 
By contrast, P3's easier task T4 (\textit{relatively easy}, 2/5 difficulty, 3/10 effort) had between 10-15 invocations per quartile.
For Verus, two tasks (T3 for P7 and T2 for P5) showed steep increases in invocations toward the end, consistently exceeding 30 per quartile.
In both of these attempts, participants ran out of time and did not successfully complete the task.
In contrast, we see that the same tasks were successfully completed by P8 (T3 and T2), with fewer verifier invocations -- at most half as many per quartile. 
P8 reported a 3/5 difficulty and 4/10 effort for T3, and 2/5 difficulty and 3/10 effort for T2. This indicates a positive relationship between the self-reported effort in a task and the number of invocations at each quartile.
P8 was successful in completing both assigned tasks with far fewer verifier invocations overall.
 
\subsubsection{Frequency of Verification Errors}




Overall, the average number of errors active at any given time across all tasks was 1.49 for F* and 1.9 for Verus. However, we observe wide variability in the distribution of active errors across tasks, with some participants employing a disciplined approach to keep the number of active errors low, and others allowing the number of errors to spike to 35-40 in Verus. 
For example, for both of their tasks, P8 intentionally maintained 0-4 active errors across tasks, achieving better outcomes than peers on the same tasks, who encountered more than 5 or even more than 15 active errors at several points throughout the task. 
This suggests a potential association between disciplined strategies to purposefully keep error counts low (by deferring verification on segments of the code and tackling the proof section by section) and successful task outcomes.
The nature of the errors encountered by users also varied by task. For example, in some cases errors were predominately related to syntax while other cases had verification errors relating particularly to asserts and invariants, e.g. P7's T1 (9.3\% invariant), P8's T3 (assertions 23.7\%). 


\section{RESULTS: Emergent Strategies }
  \label{sec:qual}

\subsection{RQ3a: Strategy Themes from Thematic Analysis of Post-Task Questionnaires}
  \label{sec:rq3a}
We conduct a qualitative Thematic Analysis of expert self-reports about the development process, collected through the post-task survey, we consolidate them as a set of high-level themes:
\subsubsection{Experts Emphasize Task Decomposition and Iterative Refinement}\label{rq3:planning} 
A predominant theme is the difficulty involved in reasoning about the correct specification.
Participants mitigate this by decomposing a task into smaller, manageable subgoals, each with a well-defined purpose. This includes decomposition of the implementation as well as specifications.
Workflows typically began with high-level specifications, adding pre- then postconditions before addressing details. 
For example, P1 broke down task T3 into two subgoals -- deduplication and intersection, which allowed for focused development of separate specification and functions for each subgoal, followed by implementation of each subgoal independently.  
P8 succinctly describes this approach: "Start with the requires/ensures, then add assume(false), then add invariants and asserts.  When stuck, search for the underlying cause of the failure by adding more precise assume and assert.".
This step-by-step decomposition also made it easier to iteratively refine subgoals, rather than working to solve the task in top-down or bottom up manner.

For example, participants explain that it is easier to iteratively revisit specifications to strengthen them after implementing portions of the functionality, rather than aiming to draft the full specification upfront. Decomposition is also crucial when tackling complex cases that are harder to reason about.
Common sources of difficulty mitigated through decomposition include: proving correctness with existential quantifiers, reasoning about loops or sequence iteration, or edge cases like integer overflow. 
%
%


\subsubsection{Experts Define Specs Early, and Align Spec Structure with the Implementation}
%
%
Experts prioritized drafting specifications early in the task.
In F*, they also deliberated on the most suitable representation for the implementation, balancing efficiency with proof complexity—e.g., choosing between \texttt{Seq} and \texttt{List} for T3.
Considering which quantifiers to use also impacted task difficulty. For example, P8 notes: "I spent most of the time on what seemed like a simple property of subranges but eventually decided to rewrite the invariant directly using "for all", not subranges, and the "for all" version worked immediately."
Decisions on quantifiers also influenced task difficulty. P8 shared: "I spent most of the time on what seemed like a simple property of subranges but eventually decided to rewrite the invariant directly using "for all", not subranges, and the "for all" version worked immediately."
Telemetry data confirmed that experts emphasized aligning specification structure with implementation, as P1 put it: "For the specification code I tried to match the structure of the implementation since I've found that that generally makes the proof easier". Similarly, P5 said that one should "Set up convenient `spec' functions for each portion of the task, such that they lend themselves to more natural proof w.r.t direction of iteration."
%

\subsubsection{Experts are Familiar with External Libraries, Select and Modify Lemmas As Needed}\label{sec:rag}
Experts emphasized the importance of finding the relevant lemmas before diving into the proof process, but their approaches to using library definitions varied.
Several chose to not consult documentation due to familiarity and lack of time during the study. 
Some reused existing library definitions as-is, while others modified or re-implemented them based on familiarity and task requirements. For example, strengthening existing lemmas if their specifications are too weak,  P1 said: “I tried to use the filter from FStar.List.Tot, but its spec was not strong enough, so I rewrote it with a tighter specification.”

\subsubsection{Experts Use Measured and Frequent Feedback from the Verifier}\label{sec:verifier}
Experts relied heavily on verifier feedback, using error messages and visual cues to guide each development step.
Several intentionally hide away error messages through \texttt{assumes, admits} or comments to make the `proof go through', and  iteratively address statements to make specifications and invariants verifiable. 
P2 and P4 mentioned testing frequently with assertions and iterating using feedback from the verifier.
P2 also added test cases explicitly (e.g., assert\_norm) to catch incorrect assumptions.
A recurring theme was the need for clearer error messages, with P4 highlighting frustration over "opacity in what's missing and errors."

\begin{table*}[t!]
\centering
\caption{Per‑session behavioral features used in RQ3b strategy analysis.
\emph{Q1}= first 25\% of session time. }
\label{tab:features}
\begin{tabular}{@{}llp{6.2cm}@{}}
\toprule
\textbf{Feature} & \textbf{Definition} & \textbf{Intent} \\ \midrule
$\mathit{early\_spec}$ &
$\dfrac{\#\text{SPEC edits in Q1}}{\#\text{all edits in Q1}}$ &
Front‑loaded specification effort. \\[6pt]

$\mathit{verify\_fq}$ &
$\dfrac{\#\text{verifier invocations}}{\text{task duration}}$ &
Aggressiveness of solver invocation. \\[8pt]

$\mathit{clean\_state}$ &
$\dfrac{\text{time with }\,\text{active\_errors}=0}{\text{task duration}}$ &
Discipline in keeping active errors low \\[12pt]

$\mathit{pause\_frac}$ &
$\dfrac{\text{time gaps}>5\text{s}}{\text{session duration}}$ &
Proxy for deliberate “think time.” \\[8pt]

$\mathit{defer}$ &
$\dfrac{states\ with \ DEFER }{\text{all states}}$  &
The percentage of time in a state with  \texttt{assume} or comments to defer sub‑goals. \\
\bottomrule
\end{tabular}
\end{table*}
\subsection{RQ3b: Telemetry-Derived Strategies and Task Outcomes}

To ground the qualitative themes surfaced in (Section~\ref{sec:rq3a}), we apply an
unsupervised analysis to cluster whole sessions using five behavioural
signals mined from the telemetry (Table~\ref{tab:features}).  Each
\emph{session} is a (participant, task) pair.  Events are partitioned into
quartiles of elapsed time; the first quartile (Q1) captures “early’’
behaviour.  Feature values are $z$‑normalised \emph{within each
language} before clustering to avoid language‑specific scale effects.

\paragraph{Feature Selection.}
The selected features are directly informed by the results of RQ1, RQ2. All clusters use the common planning / interaction features: 1)
$\mathit{early\_spec}$ the proportion of Q1 edits that are related to specification drafting, 2) the frequency of verifier invocations across the task $\mathit{verify\_fq}$,  3) the amount of time spent in paused states $\mathit{pause\_frac}$, 4) the proportion of task time with no active verification errors $\mathit{clean\_state}$, and 5) $\mathit{defer}$ the percentage of time in a state with deferred subgoals. (Table~\ref{tab:features}).  

\paragraph{Clustering method.}
We run $k$-means (scikit-learn, $n_{\text{init}}{=}\texttt{auto}$, seed~42)
\emph{separately} for F$^{\!*}$ and Verus, evaluating $k\!\in\!\{2,3,4\}$.
For F*, $k{=}3$ offers the best balance of separation and interpretability
(silhouette $0.34$; $k{=}2$: $0.29$, $k{=}4$: $0.21$) and yields
Planner/Rapid/Balanced groups (sizes 2/3/4).
For Verus, $k{=}3$ attains the highest silhouette ($0.173$; $k{=}2$: $0.153$,
$k{=}4$: $0.092$) but produces an unbalanced 1/2/5 cluster sizes with a
singleton in the spec-first planner cluster. We retain the $k{=}3$ grouping for descriptive
completeness while treating comparisons involving the singleton as
exploratory. 

Results of K-means clustering for F$^{\!*}$ are shown in
Table~\ref{tab:cluster-outcomes-fstar}.  Three archetypes emerge:

\begin{itemize}[leftmargin=1.6em]
  \item \textbf{Spec‑first Planners ($C_S$)}: highest early specification
        investment (median $74\%$ Q1 edits) and the largest
        $\mathit{clean\_state}$ (frequent returns to a zero‑error state)
        with moderate verifier invocations.
  \item \textbf{Rapid Verifiers ($C_R$)}: lowest early specification drafting ($\approx46\%$) but
        very high relative verifier frequency ($\sim7.3$ calls/min) coupled with
        near‑zero non-error states.
  \item \textbf{Balanced ($C_B$)}: mid‑range early specification investment ($\approx$ 50\%), lowest
        invocation rate, and intermediate time  in clean non-error state.
\end{itemize}

In general, Planners achieve the shortest median completion time and perfect
success rate, Rapid sessions are slower and less successful, and the Balanced
sessions fall in between both Rapid and Planners.


\begin{table*}[]
\centering
\caption{F$^{\!*}$ cluster centroids with outcome metrics.}
\label{tab:cluster-outcomes-fstar}
\begin{tabular}{@{}lccccccc@{}}
\toprule
 & $\mathit{early\_spec}$ & $\mathit{verify\_fq}$ &
   $\mathit{clean\_state}$ & $\mathit{pause\_frac}$ & $\mathit{defer}$ &
   Success & Median Dur. \\ \midrule
\textbf{C\textsubscript{S}}: Spec-first Planner  &
\textbf{0.74} & 3.63 & \textbf{0.07} & 0.06 & \textbf{0.77} &
1.00 & 21.89 \\
\textbf{C\textsubscript{B}}: Balanced  &
0.50 & 1.78 & 0.04 & \textbf{0.07} & 0.28 &
1.00 & 22.74 \\
\textbf{C\textsubscript{R}}: Rapid-verify    &
0.46 & \textbf{6.54} & 0.01 & 0.03 & 0.63 &
0.67 & \textbf{24.41} \\
\bottomrule
\end{tabular}
\end{table*}

\begin{table*}[]
\centering
\caption{Verus cluster centroids with outcome metrics. Success rate = fraction of sessions
in the cluster that completed; median duration in minutes.}
\label{tab:verus-centroids}
\begin{tabular}{@{}lccccccc@{}}
\toprule
 & $\mathit{early\_spec}$ & $\mathit{verify\_fq}$ &
   $\mathit{clean\_state}$ & $\mathit{pause\_frac}$ &
   $\mathit{defer}$ & Success & Median Dur. \\ \midrule
\textbf{C\textsubscript{S}}: Spec-first Planner &
\textbf{0.56} & 3.74 & \textbf{0.57} & 0.02 & 0.00 &
\textbf{1.00} & 11.77 \\[3pt]
\textbf{C\textsubscript{B}}: Balanced &
0.53 & 2.18 & 0.20 & \textbf{0.05} & \textbf{0.19} &
\textbf{1.00} & 25.12 \\[3pt]
\textbf{C\textsubscript{R}}: Rapid-verify &
0.44 & \textbf{5.52} & 0.10 & 0.03 & 0.00 &
0.60 & \textbf{26.10} \\
\bottomrule
\end{tabular}
\end{table*}

Clustering the Verus sessions ($k{=}3$) yields one singleton and two
multi–session groups (Table~\ref{tab:verus-centroids}). The singleton
\emph{specification first planner} ($C_S$) exhibits the highest early specification share
(0.56) and an exceptionally high clean\_state (0.57), finishing in
11.8\,min with success. Pariticpants in the \emph{Balanced} cluster ($C_B$; $n{=}2$)
also front load specification drafting (0.53) but proceed with the lowest early
verifier cadence (2.18/min in Q1), and have the most think pauses (pause\_frac
0.05) and comment/assume edits (0.19) to decompose the task into sub–goals. Both of the Balanced sessions
succeeded (median 25.1\,min). The \emph{Rapid verifiers} ($C_R$; $n{=}5$)
draft less specification early (0.44) while invoking the verifier most aggressively
(5.52/min), in addition, they achieve the lowest clean\_state (0.10) and only 60\%
success, with the widest duration variability (IQR 32.2\,min).

Across both languages, an effective proof strategy, i.e. one that leads to successful and time-efficient task completion, is characterized
by (i) deliberate early specification drafting, (ii) measured
verifier invocation, and (iii) disciplined error management with frequent
returns to a clean state. In F$^{\!*}$ we distinguish three styles—\textsc{Planner},
\textsc{Rapid}, and \textsc{Balanced}. Planners combine the highest
early\_spec and clean\_state with strong use of defer mechanisms; Rapid-verification
sessions substitute extreme verifier invocation for upfront spec (lower
success / longer median time), while Balanced sessions eventually
succeed but incur higher variability (large IQR). In Verus, after enforcing a minimum cluster size, the data
collapse to a robust two‑way split: \textsc{Planners} (higher early spec,
cleaner frontier) versus \textsc{Rapid} verifiers (lower early spec,
higher early polling, persistently error states).  

While the small sample of sessions limits
statistical significance and may mask rarer strategies, the
independent emergence of distinct clusters in two languages and the
consistent impact on task outcome indicates that these
patterns may scale. 

\subsection{RQ3b: Case Study of Strategy Archetypes}

\begin{figure*}[th!]
  \centering
  \begin{minipage}{0.8\textwidth}
    \includegraphics[width=\linewidth]{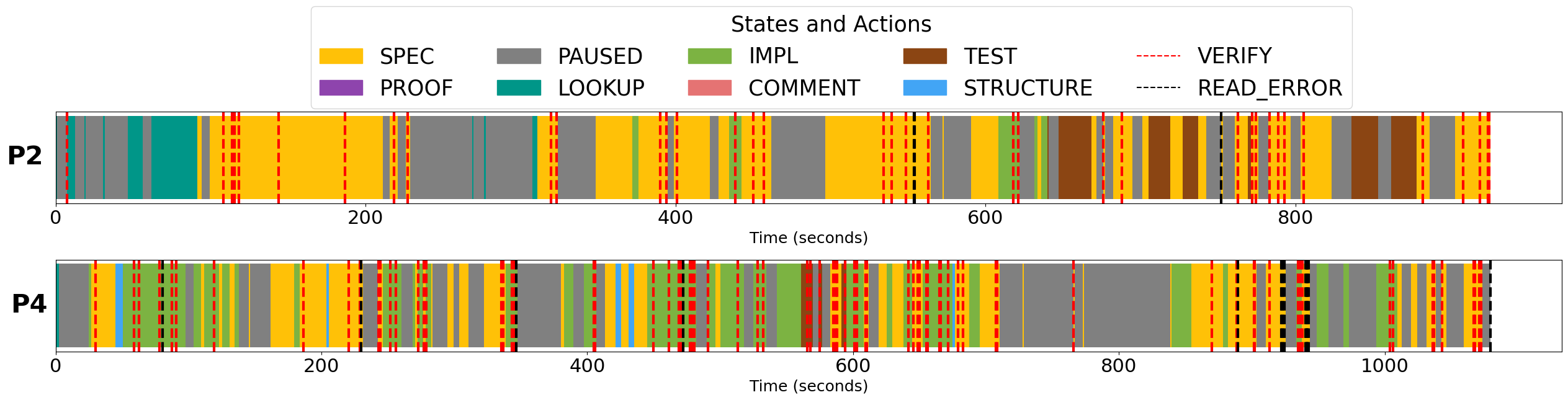}
  \end{minipage}
  \hspace{1em}
  \begin{minipage}{0.8\textwidth}
    \includegraphics[width=\linewidth]{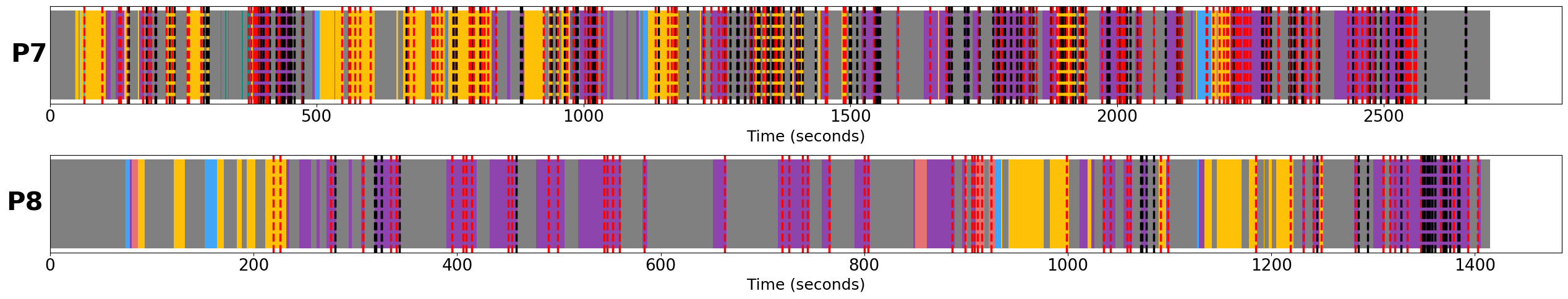}
  \end{minipage}
  \caption{States and actions over time during task T3 for F* participants P2 (\textbf{C\textsubscript{S}}) and P4 (\textbf{C\textsubscript{B}}), and Verus participants P7 (\textbf{C\textsubscript{R}}) and P8 (\textbf{C\textsubscript{B}}).}
  \label{fstar-249}
\end{figure*}

To gain deeper insight into how the three strategy archetypes are surfaced in practice, we use Task T3 as a case study to illustrate participant behaviors associated with each strategy.

\subsubsection{Case study– task T\textsubscript{3}: four timelines, three strategies}

Figure~\ref{fstar-249} overlays proof–writing \emph{states} (coloured
blocks) and \emph{actions} (dotted lines) for all four T\textsubscript{3}
sessions.  The visualization highlights how successful and unsuccessful
approaches unfold across languages.

\paragraph{F*: two paths to success.}
Both \textbf{P2} and \textbf{P4} eventually implemented binary search and verified the specification, yet
their trajectories differ markedly:

\begin{itemize}[leftmargin=1.8em]
  \item \textbf{P2 – \textbf{C\textsubscript{S}} specification-first planner}  After an initial 
        pause to plan,
        P2 performs lemma look‑ups,
        drafts the high‑level specification,
        and then iteratively alternates between
        \textbf{IMPL} and \textbf{SPEC} states
        while recycling existing lemmas for deduplication.
        Dedicated \textbf{TEST} edits appear only once
        the main proof writing stabilizes.
        The entire solution fits in \emph{30} lines and was completed in
        $\approx$14 min.

  \item \textbf{P4 – \textbf{C\textsubscript{B}} Balanced approach.}  P4 writes \emph{all}
        multiple supporting lemmas about sorted lists from scratch, and switches rapidly between
        \textbf{SPEC}, \textbf{IMPL},
        and \textbf{PROOF} states.
        This solution is \emph{97} lines (over double) and added 
        $\approx$5 min in task time.
\end{itemize}

\noindent
One of the primary differences between both tasks is that P4 wrote multiple lemmas about sorted lists to prove that the resulting list contained no duplicates, whereas P2 opted to use a function to remove duplicates.  Additionally, P4 frequently switched between developing specifications and implementation, while P2 worked in distinct phases, including dedicated testing. For this particular task, leveraging existing libraries not only shortened the final code and proof
(3× fewer lines) but also reduced the time on task.

\paragraph{Verus: one successful and one unsuccessful outcome.}
In Verus, only \textbf{P8} completed the task:

\begin{itemize}[leftmargin=1.8em]
  \item \textbf{P7 – \textbf{C\textsubscript{R}} frequent verification calls.}
        The timeline is dominated by dense
        \textbf{VERIFY} invocations and repeated
        \textbf{SPEC} rewrites.  Despite 45 min of effort,
        the participant finishes the task with an incomplete proof and implementation.
        
  \item \textbf{P8 –\textbf{C\textsubscript{B}} Balanced/Specification first planner.}
        P8 sketches the full \textbf{SPEC} \& proof skeleton
        first, inserts temporary \texttt{assume} statements, and then
        resolves sub‑goals one by one.  Notice the \textbf{grey}
        thinking pauses \emph{after} each verifier invocation. Although clustered as $C_B$, for Verus $C_B$ and $C_S$ clusters show little variance. 
\end{itemize}

\noindent

Excessive verifier invocation coupled with spec churn (P7) contrasts sharply with P8’s plan‑then‑verify approach, reinforcing the importance of keeping active errors low and deliberate pauses to think.
The two most successful sessions (P2, P8) share these traits: 1) high initial investment in the specification, 2) reuse of trusted lemmas 3) incremental implementation of commented‑out post‑conditions as sub goals, and 4)
explicit thinking pauses.

These patterns echo the \emph{spec‑first planner} strategy quantified earlier, suggesting that effective proof agents should aim to imitate these strategies and be mindful of getting stuck in frequent verification loops or tolerating a large number of active errors.

\begin{figure*}[h!]
  \centering
  \begin{minipage}{\textwidth}
  \centering
    \includegraphics[width=0.9\linewidth]{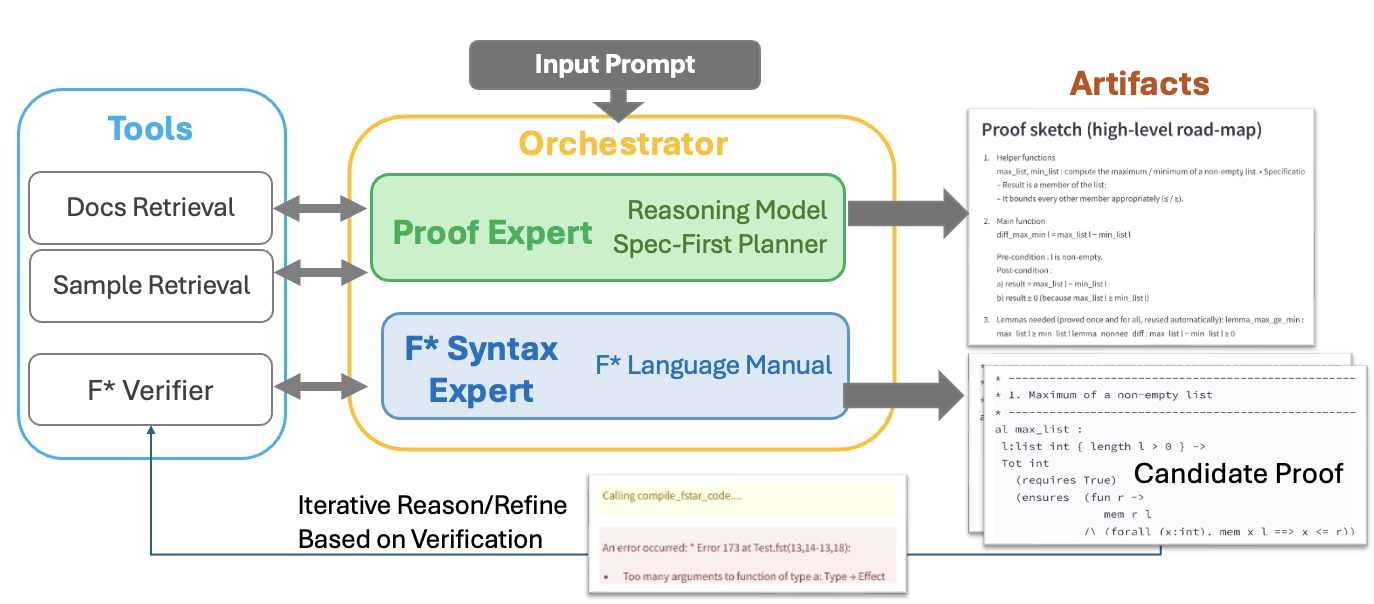}
  \end{minipage}
  \caption{A two-agent system where first the Proof Expert Agent generates a proof sketch and outlines a high-level roadmap: decomposing the task into helper functions, defining specifications, and identifying lemmas. The sketch is passed to the F* Syntax Expert Agent, which produces syntactically valid F* code (as shown in the code box at the right) guided by verifier feedback}
  \label{fig:agent}
\end{figure*}
\section{RECOMMENDATIONS}
We propose design recommendations for an AI-assisted proof copilot that can improve productivity, reduce cognitive load, and improve flow during proof development.

\subsection{Proof Copilot Design}
\subsubsection{Online Lemma and Documentation Retrieval}
The copilot should support quick retrieval of relevant lemmas from libraries or example code snippets.
These suggestions should be tailored to the current proof state and subgoals. 
When multiple representations are possible, the lemma suggestions themselves can guide users toward stronger or more appropriate specifications (\autoref{sec:rag}).

\subsubsection{Task Planning and Decomposition}
Rather than jumping into code or proofs, the copilot should offer step-by-step suggestions: which helper functions to write, how to organize the specification, and what sequence of sub tasks to follow (\autoref{rq3:planning}).
Since many users begin with high-level specifications, the copilot can assist by turning these into partial proof sketches.
Given a clear pre/post condition, the copilot could outline a plan—suggesting what needs to be shown, where helper lemmas might fit, and what refinement holes to fill.
In addition, when users are stuck in tight edit-verify-edit loops, they are likely not in a productive state.
The copilot should recognize these patterns and intervene only then, and precisely offer suggestions or decompositions to reorient the user.

\subsubsection{Proof Debugging and Targeted Verifier Feedback}

One opportunity is helping users make sense of verifier feedback. Instead of showing long, dense error messages, the copilot could act as an interpreter—explaining likely causes, pointing to the proof steps involved, and suggesting concrete changes.
Instead of relying on users to comment out failing code or manually add defer, a better user experience would be breakpoint-like toggles to ignore specific lines during verification. A model trained on common proof mistakes could also identify which errors to prioritize or hide away for the user. 
This would reduce time spent guessing which part of the proof to fix and encourage faster iteration (\autoref{sec:verifier}).

\subsubsection{Agent Specialization and Modularity}

Rather than building one assistant, we suggest creating specialized agents for different proof tasks: lemma search, proof sketching, syntax expert, verifier feedback interpretation and also add appropriate tool calling capabilities in these agents~\cite{masterman2024landscape}.
This modular approach would let each agent be tuned for a specific job, while the orchestrator and/or the proof expert stays in charge of high-level creative decisions.
The next section details one such agentic system for F* proofs we implemented in the Autogen framework~\cite{wu2024autogen}.

\subsection{F* Agent Case Study}

Inspired by patterns surfaced in RQ3, we propose a proof agent system design for F* and explore how a multi-agent architecture can emulate expert behaviors.
We implement this proof-of-concept agent using AutoGen~\cite{wu2024autogen}.
Both specification-first planning strategies and themes from participants self-reports emphasized the importance of tasks decomposition into smaller components and iterative refinement of proof and specification. Our two-agent system mirrors this iterative workflow.

The F* agent as shown in \autoref{fig:agent} uses GraphRAG \cite{edge2024local} to retrieve relevant documentation and examples from the F* documentation and language guide. This mimics the expert's inherent knowledge of useful F* libraries and lemmas.
The proof expert is instructed to act as a specification-first planner and generates a high-level proof sketch, thinking through: 1) the specifications 2) how to decompose the problem 3) which representations to use 4) how to align the specification with the implementation, and 5) how to organize the overall proof strategy. We use the state-of-the-art reasoning model, OpenAI's o3~\cite{openai_o3_2025}, as the underlying proof model. 
The F* syntax expert then takes over, and translates the plan into well-formed F* code while handling syntax-level issues, using its access to the F* language manual. This separation is informed by the observation that experts frequently iterated between writing specs and fixing low-level issues within the context of a high-level plan.
The code generated by the syntax agent is passed to the F* verifier. The example in \autoref{fig:agent} highlights a verifier failure, shown in red, due to a type mismatch.
The bottom of the figure shows a feedback loop, where verification results are interpreted and used to refine the code or sketch, continuing the process iteratively, just like an expert would debug.

\subsubsection{F* Agent Evaluation}
We evaluate our agent on tasks T4 and T1, which were rated by expert users as the easiest and most difficult tasks, respectively, in the user study.
The F* agent produced correct, verified solutions with 8 (T4) and 7 (T1) refinement loops, as counted by the number of verifier invocations. 
Here we report only *effective* turns (those that invoke the F$^\star$ verifier), though internal reasoning and inter-agent conversations are not included in this refinement count. 
Interestingly, the harder task (T1) required no additional refinement loops over T4. We observe that (i) the proof agent’s planning phase generates high quality initial sketches and relevant lemmas for complex specifications, and (ii) the verifier’s feedback is helpful to providing guidance on both syntax and proof failures. 

We compare the F* proof agent to a naive baseline,  using OpenAI's state-of-the-art  \texttt{o4-mini}~\cite{openai_o3_2025} as the underlying LLM. We maintain the same system and high level task prompts, without instructions for agent tool use. For the selected tasks, the baseline approach was unable to achieve a correct verified solution within 30 verifier refinement loops,  indicating a potential $ 3.75\times$ reduction in verification calls for our agent versus an unconstrained LLM attempting iterative self‑correction.

\section{THREATS TO VALIDITY}

Instrumentation bugs or mis-annotation of events could impact the results. We mitigated this by (i) cross–checking a samples of
telemetry segments against screen recordings, (ii) deriving all behavioural features from automatically tagged states rather than manual coding, and
(iii) re–running the k-means clustering after alternative normalizations to confirm stability of the qualitative
labels. 
The limited number and demographic of user study participants may not be representative and derived strategies may not represent novice or intermediate users. In addition, small sample size constrains statistical significance and may obscure less common strategies. However, the independent emergence of distinct strategy clusters across two languages, along with consistent impacts on task outcomes, suggests that these patterns may generalize. 
The selected tasks are small to medium in scope and may not capture
maintenance activities (proof refactoring,
multi–module reasoning) typical of tasks in real-world verified systems. 

Threats related to our Thematic Analysis include threats related to participant subjective self‑reports and researcher bias in coding. Threats related to the proof agent case study include the small sample size, and limited sampling seeds. Reported gains may not transfer to more complex tasks or different languages and evaluation on broader benchmarks is part of future work.

\section{CONCLUSION}

This work offers one of the first fine‑grained, cross‑language studies about how experts actually engineer proofs in proof‑oriented
programming languages, F* and Verus. The collected telemetry shows three distinct task trajectories, which we present as
 strategies (specification-first planning vs.\ rapid verification behaviors) and
informal practices (deliberate pauses, selective sub-goal deferral) that are
not captured in final code snapshots but predictive of task outcomes. This work highlights that formal proofs often depend on abundant informal information and intermediate reasoning steps, which do not appear in the final code but may be crucial signals for future model and tool design. We suggest concrete agent design opportunities: assistants should bias toward
early spec drafting, sub-goal decomposition, bounded active errors, and maintain disciplined interaction with the verifier. We plan to compare our findings across expertise levels, with additional participants and representative tasks, as well as further leverage telemetry data in future work.

\balance

\bibliographystyle{ACM-Reference-Format}
\bibliography{main}

\balance
\end{document}